\documentclass{PoS}

\title{Capabilities of the CMS detector for studies of hard probes in heavy ion 
collisions at the LHC}

\ShortTitle{Capabilities of the CMS detector for studies of hard probes in heavy 
ion collisions at the LHC}

\author{\speaker{Igor Lokhtin} (for the CMS Collaboration)\\
        D.V. Skobeltsyn Institute of Nuclear Physics, M.V. Lomonosov Moscow
	State University, Moscow, Russia \\
        E-mail: \email{Igor.Lokhtin@cern.ch}}

\abstract{The capabilities of the CMS experiment to study properties of hot and dense
QCD-matter created in heavy ion collisions at the CERN Large Hadron Collider with the 
perturbative processes (so-called ''hard probes'') are presented. Detailed
studies from complete simulations of the CMS detectors in Pb+Pb collisions at $\sqrt{s}=5.5$ TeV per
nucleon pair are presented in view of two hard probes: quarkonium and 
$\gamma$-jet production.} 

\FullConference{High-pT Physics at LHC - Tokaj'08\\
		 March 16 - 19 2008\\
		 Tokaj, Hungary}

\begin{document}

\section{Introduction}
The study of the fundamental theory of the strong interaction (Quantum Chromodynamics, 
QCD) in new, unexplored extreme regimes of super-high densities and temperatures is one
of the primary goals of the modern high energy physics. The experimental and 
phenomenological study of multiple particle production in ultrarelativistic 
heavy ion collisions is expected to provide valuable information on the 
(thermo)dynamical behaviour of QCD matter in the form of a quark-gluon plasma (QGP), 
as predicted by lattice QCD calculation. A detailed description of the potential of CMS to carry out a series of 
representative Pb-Pb measurements has been presented in~\cite{D'Enterria:2007xr}. 

Heavy ion observables accessible to measurement with CMS include:
\begin{itemize}
\item ``Soft'' probes~\cite{ferenc}: global particle and energy rapidity 
densities, elliptic flow and spectra 
of low transverse momentum hadrons. These observables are mostly sensitive to the 
space-time evolution of the system once thermalization has set in and thus carry 
information about the thermodynamical properties of the produced QCD matter.
\item ``Hard'' probes: quarkonia, heavy quarks, jets, $\gamma$-jet
and high-$p_{T}$ hadrons, which are produced with high transverse momenta $p_T$ or 
large masses $M$ (much greater than the typical QCD scale of confinement: $p_T$, $M$ 
$\gg$ $\Lambda_{\rm QCD}=200$ MeV). The hard probes production cross sections can be 
described in the framework of perturbative QCD theory. Such hard particles are 
produced in the very early stages of the evolution of the system and thus are 
potentially affected by final-state interactions as they traverse the produced medium. 
Modifications with respect to the ``vacuum QCD'' spectra and cross-sections measured in 
proton-proton collisions, provide direct information on the dynamical and transport 
properties of the system: initial parton densities, transport coefficient of the 
medium, critical energy density. 
\end{itemize}

In this contribution, the detailed study for a complete simulation of the CMS detectors 
in Pb+Pb collisions at $\sqrt{s}=5.5$ TeV per nucleon pair are presented in view of 
two hard probes: quarkonium and $\gamma$-jet production. The results of similar studies
for high-$p_{T}$ hadrons are presented during this Workshop in another
talk~\cite{krisztian}. 

\section{CMS detector}
CMS is a general purpose experiment at the LHC designed to explore the physics at the 
TeV energy scale~\cite{ptdr}. Since the CMS detector subsystems have been designed 
with a resolution and granularity adapted to cope with the extremely high luminosities 
expected in the proton-proton running mode, CMS can also deal with the large particle 
multiplicities anticipated for heavy-ion collisions. A detailed description of the 
detector elements can be found in the corresponding Technical Design 
Reports~\cite{Htdr,Mtdr,Etdr,Ttdr}. The central element of CMS is the magnet, a 13~m 
long, 6~m diameter, high-field solenoid (a uniform 4 T field) with an internal radius 
of $\approx 3$ m.

The tracker covers the pseudorapidity region $|\eta|<2.5$ and is composed of two 
different types of detectors: silicon pixels and silicon strips. 
The pixel detector consists of three barrel layers located at 4, 7, 11 cm from the
beam axis with granularity $150 \times 150$ $\mu$m$^2$ and two forward layers with 
granularity $150 \times 300$ $\mu$m$^2$ located at the distances of 34 and 43 cm in 
z-direction from the center of detector. Silicon strip detectors are divided into inner and outer 
sections and fill the tracker area from 20 cm to 110 cm (10 layers) in the transverse 
direction and up to 260 cm (12 layers) in longitudinal direction. 

The hadronic (HCAL) and electromagnetic (ECAL) calorimeters are located inside the coil 
(except the forward calorimeter) and cover (including the forward calorimeter) from 
$-$5.2 to 5.2 in pseudorapidity. The HF calorimeter covers the region $3<|\eta|<5.2$.

The CMS muon stations cover the pseudorapidity region $|\eta|<2.4$ and consist of drift 
tube chambers (DT) in the barrel region (MB), $|\eta|<1.2$, cathode strip chambers (CSCs) 
in the endcap regions (ME), $0.9<|\eta|<2.4$, and resistive plate chambers (RPCs) in 
both barrel and endcaps, for $|\eta|<2.1$. 
The RPC detector is dedicated to triggering, while the DT and CSC detectors, used for precise 
momentum measurements, also have the capability to self-trigger up to 
$|\eta|<2.1$. 

\begin{figure}[!Hhtb]
  \centering
\includegraphics[width=12.cm,height=6.cm]{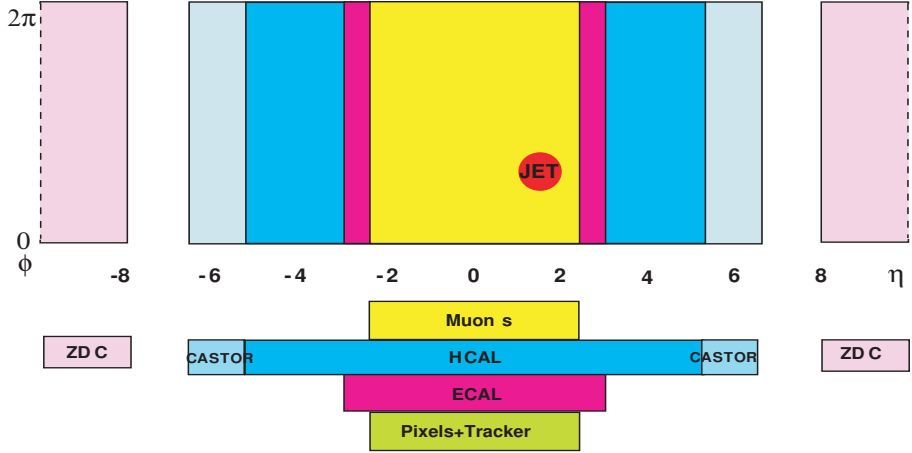}
\caption{CMS coverage for tracking, calorimetry, and muon identification in
pseudo-rapidity ($\eta$) and azimuth ($\phi$). The size of a jet with cone $R=0.5$
is also depicted for comparison.} %illustration.}
\label{fig:cms_accept}
%\bigskip
\end{figure}

Note that CMS is the largest acceptance detector at the LHC (Fig.~\ref{fig:cms_accept}) with 
unique detection capabilities also in the very forward hemisphere with the CASTOR 
(5.1 $<|\eta|<$ 6.6) and the Zero-Degree (ZDCs, $|\eta_{neut}|>$ 8.3)
calorimeters~\cite{fwd_cms}. 

Another key aspect of the CMS hard probe capabilities for heavy ion physics is its 
unparalleled high-level-trigger (HLT) system running on a filter farm with an 
equivalent of ${\cal O}$(10$^4$) 1.8~GHz CPU units, yielding few tens of 
Tflops~\cite{groland}. The 
HLT system is powerful enough to run  ``offline'' algorithms on every single 
Pb+Pb event delivered by the level-1 trigger, and select the interesting events while 
reducing the data stream from an average 3~kHz L1 input/output event rate down to 
10--100 Hz written to permanent storage. The resulting enhanced statistical reach for 
hard probes is a factor of $\times$20 to $\times$300 larger, depending on the signal, 
than for the min-bias (MB) trigger.  

\section{Simulation and analysis of J/$\psi$ and $\Upsilon$ production}

One of the important hard probes at the LHC will be the production of heavy-quark 
bound states, which should be suppressed in QGP due to colour 
screening~\cite{Matsui:1986dk}. An intriguing phenomenon is the ``anomalously'' 
strong suppression of the J/$\psi$-meson yields, observed in Pb+Pb collisions at 
SPS~\cite{Abreu:1999qw,Abreu:2000ni}. Although the interpretation of this 
phenomenon as a result of the formation of a QGP is quite plausible, 
alternative explanations have also been put forward, such as rescattering on co-moving 
hadrons. The surprisingly similar amount of J/$\psi$ suppression observed at SPS and 
RHIC energies~\cite{Adare:2006ns} is not yet fully understood. Further information 
on the nature of quarkonia suppression in hot and dense QCD-matter will 
come from the 
results at the LHC at the much higher temperatures accessible. CMS features the best 
dimuon mass resolution of any LHC detector, leading to a clean separation of the 
various quarkonia states and an improved signal over background ratio. This fact will 
open up a unique opportunity to study the threshold dissociation behaviour of the whole 
bottomonium family ($\Upsilon$, $\Upsilon`$, $\Upsilon``$) together with the charmonium 
one. Since various quarkonium states are predicted to melt at different medium 
temperatures, scan of corresponding suppression factors will serve as an effective 
QCD-matter ``thermometer''.

\begin{figure}[htbp]
\begin{center}
\resizebox{10cm}{!}
{\includegraphics{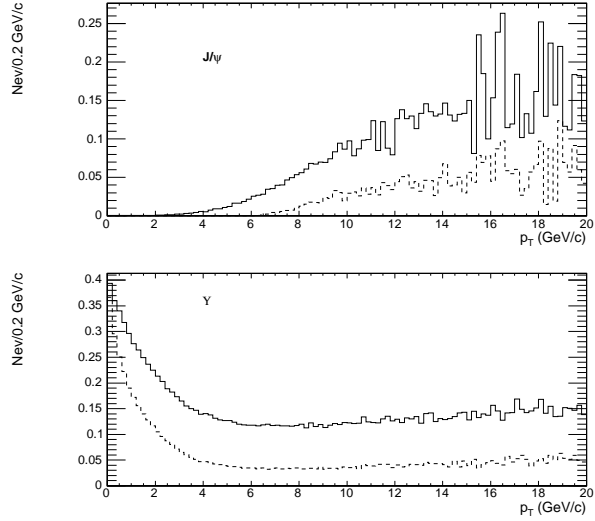}}
\caption{J/$\psi$ (top) and $\Upsilon$ (bottom) CMS acceptances (convoluted with trigger efficiencies)
as a function of $p_T$, in the full detector ($|\eta|<2.4$, solid line)
and only in the central barrel ($|\eta|<0.8$, dashed line).}
\label{fig:accept}
\end{center}
\end{figure}

Event generator HIJING \cite{hijing} with full GEANT4-based simulation of the tracking 
of secondaries and simulated detector response were 
used in the analysis presented. The J/$\psi$ and $\Upsilon$ acceptances on CMS are shown as a function of $p_T$ in Fig.~\ref{fig:accept}
for two $\eta$ ranges: full detector ($|\eta|<2.4$) and central barrel ($|\eta|<0.8$). Because of
their relatively small mass, low momentum J$/\psi$'s ($p < 4$ GeV/$c$) are mostly not accepted:
their decay muons do not have enough energy to traverse the calorimeters and coil, and are absorbed before
reaching the muon chambers. 
The J/$\psi$ acceptance increases with $p_T$, flattening out at $\sim$~15\% for $p_T>12$ GeV/$c$.
The $\Upsilon$ acceptance starts at $\sim\,$40\% at $p_T=0$~GeV/$c$ and remains constant at $\sim$~15\%
(full detector) or 5\% (barrel only) for $p_T>4$~GeV/$c$. The $p_T$-integrated acceptance is about
1.2\% for the J/$\psi$ and 26\% for the $\Upsilon$, assuming the input
theoretical distributions.

In the central barrel of the CMS detector, the dimuon reconstruction efficiency remains above 80\% 
for all multiplicities whereas the purity
decreases slightly with increasing multiplicities $dN_{ch}/d\eta$ but also stays above 80\% even at 
as high as $dN_{ch}/d\eta|_{\eta=0} = 6500$. If (at least) one of the muons is detected in the endcaps,
the efficiency and purity drop due to stronger reconstruction cuts. Nevertheless,
for the $dN_{ch}/d\eta|_{\eta=0}\approx 2000$ multiplicity realistically expected in central Pb+Pb
at LHC, the efficiency (purity) remains above 65\% (90\%) even including the endcaps.

At the $\Upsilon$ mass, the  dimuon mass resolution for muon pairs in the central barrel,
$|\eta|<0.8$, is 54~MeV/$c^2$.
In the full pseudorapidity range, the dimuon mass resolution is about 1\% of the
quarkonium mass: 35 MeV/$c^2$ at the J/$\psi$ mass and 86 MeV/$c^2$ at the $\Upsilon$ mass.
There is a slight dependence of the mass resolution on the event multiplicity.
Increasing the multiplicity from $dN/d\eta = 0$ to 2500 degrades the mass
resolution of the reconstructed $\Upsilon$  from 86 to 90 MeV/$c^2$.

\begin{figure}[!Hhtb]
\begin{center}
\resizebox{6cm}{!}
{\includegraphics{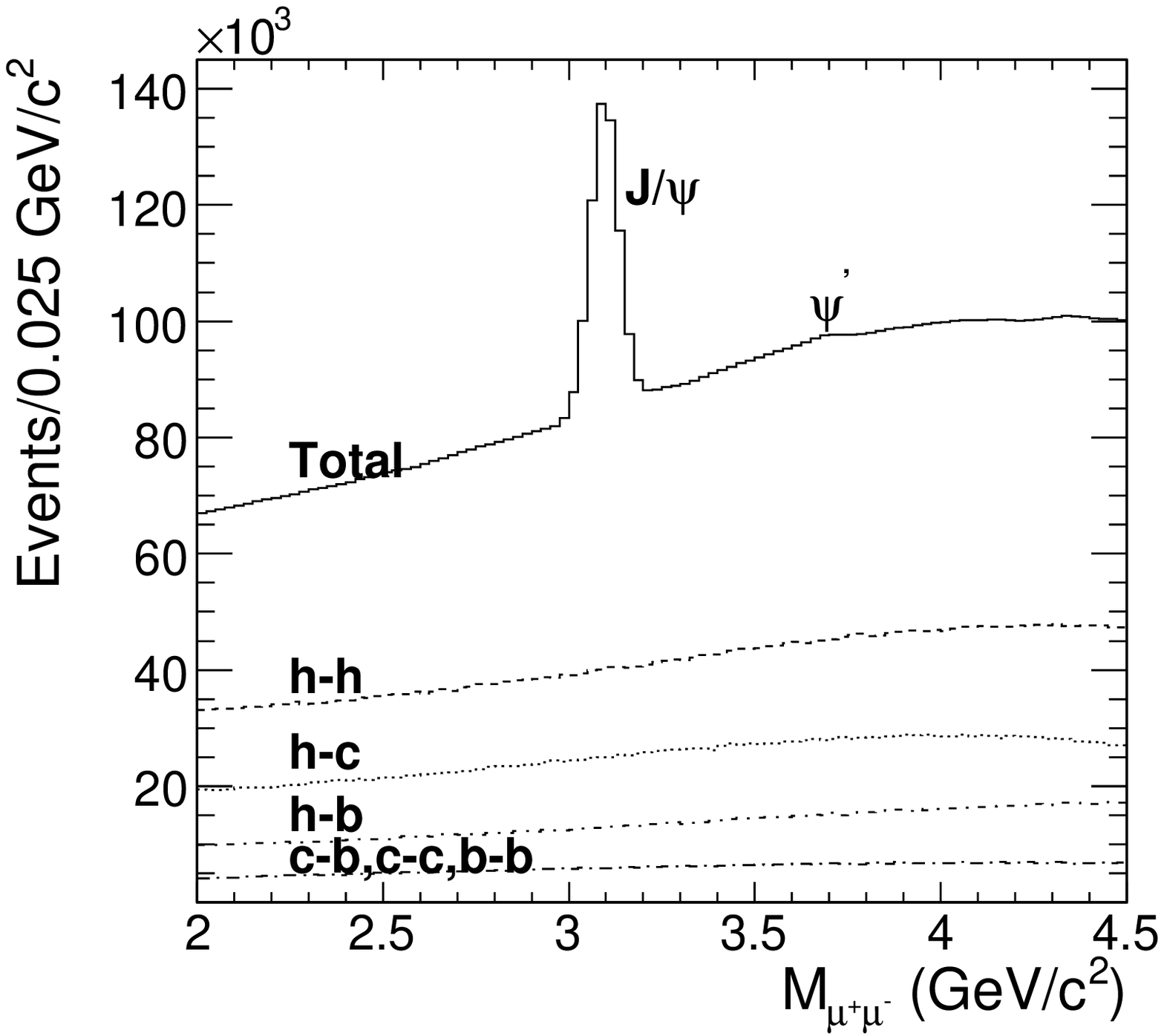}}
\hfill
\resizebox{6cm}{!}
{\includegraphics{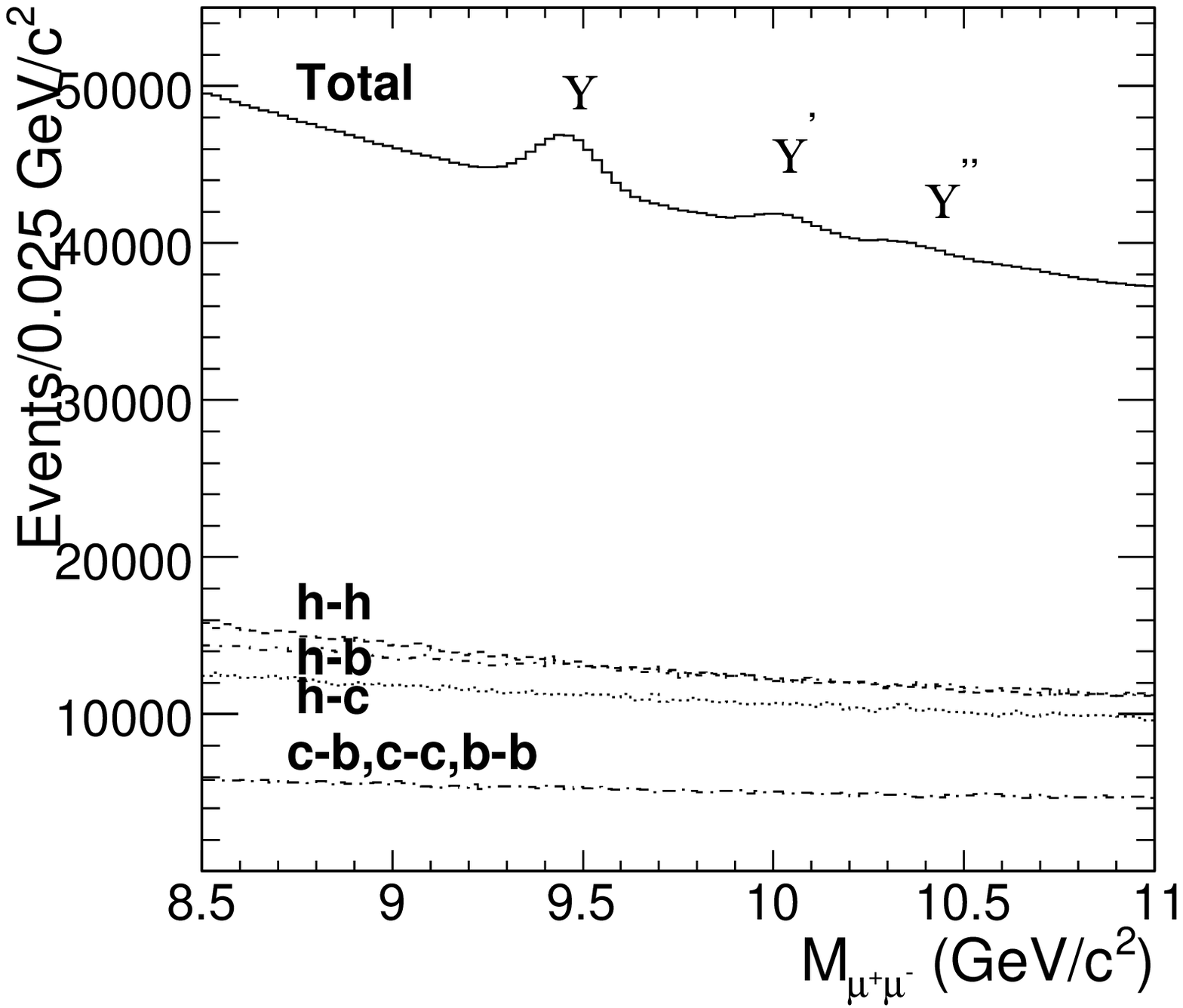}}
\resizebox{6cm}{!}
{\includegraphics{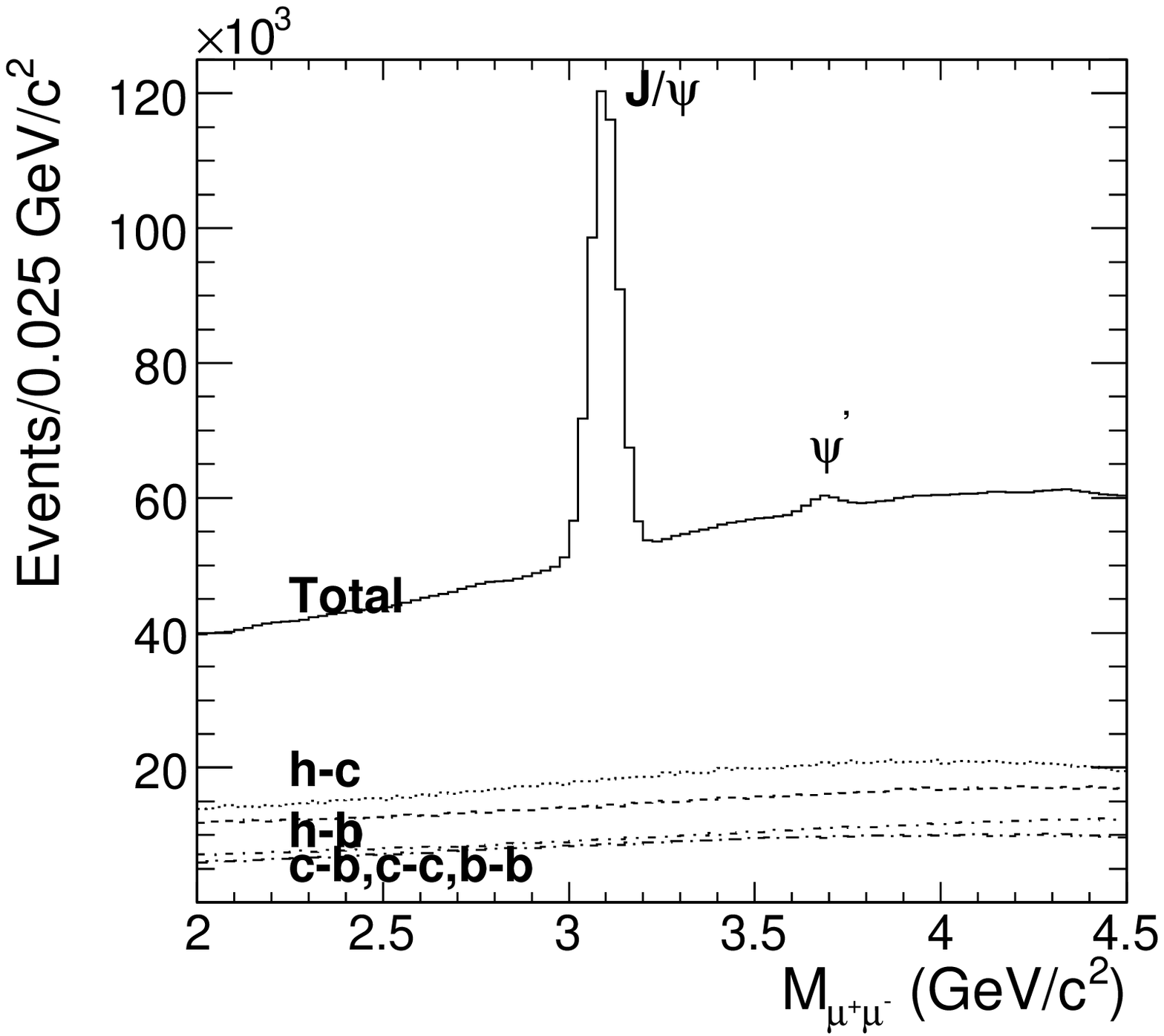}}
\hfill
\resizebox{6cm}{!}
{\includegraphics{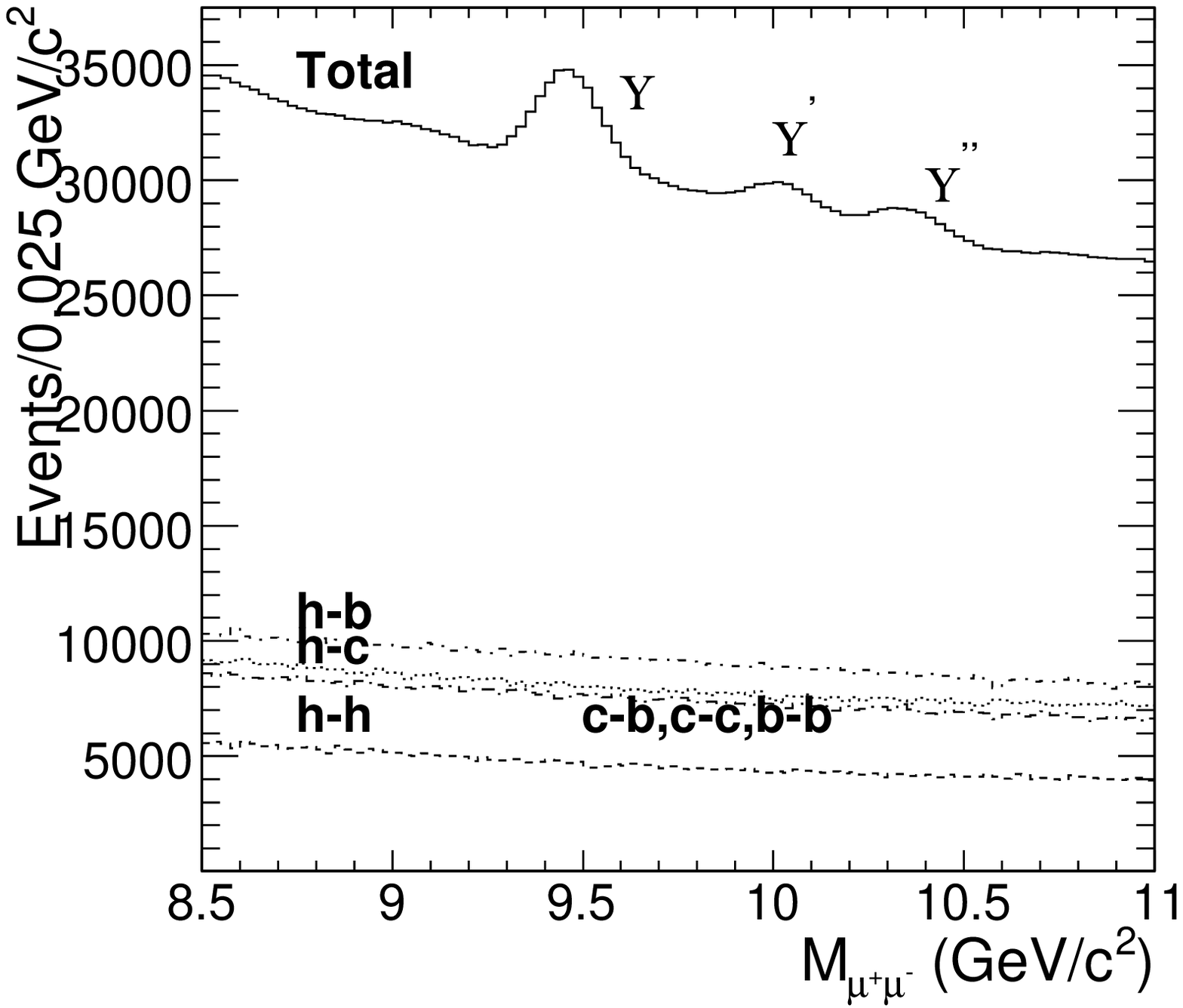}}
\caption{Dimuon mass distributions within $|\eta|<2.4$ for Pb+Pb events with
$dN_{ch}/d\eta|_{\eta=0} = 5000$ (top) 
and $dN_{ch}/d\eta|_{\eta=0} = 2500$ (bottom) in the J/$\psi$ (left) and $\Upsilon$ (right) mass regions.
The main background contributions are also shown: $h$, $c$ and $b$ stand for $\pi+K$, charm,
and bottom decay muons, respectively.}
\label{fig:minv_hm_signal_bckgds2}
\end{center}
\end{figure}

Fig.~\ref{fig:minv_hm_signal_bckgds2} shows the opposite-sign dimuon mass 
distributions, for the high and low multiplicity cases and full acceptance ($|\eta|<2.4$).
The different quarkonia resonances appear on top of a continuum due to
the various sources of decay muons: $\pi+K$, charm and bottom decays.
Assuming that the CMS trigger and acceptance conditions treat opposite-sign
and like-sign muon pairs equally, the combinatorial like-sign background can be subtracted
from the opposite-sign dimuon mass distribution, giving us a better access to the quarkonia decay
signals. The statistics of J/$\psi$ and $\Upsilon$, $\Upsilon^{'}$ and $\Upsilon^{''}$ with both muons in $|\eta|<2.4$ region 
expected in one month of data taking are 
140000, 20000, 5900 and 3500 correspondingly for the multiplicity
$dN_{ch}/d\eta|_{\eta=0}=5000$ and 180000, 25000, 7300, 4400 for
the multiplicity $dN_{ch}/d\eta|_{\eta=0}=2500$. The signal-to-back\-ground ratios are 0.6, 0.07 for J/$\psi$ and $\Upsilon$'s for 
$dN_{ch}/d\eta|_{\eta=0}=5000$ correspondingly and 1.2, 0.12 for 
$dN_{ch}/d\eta|_{\eta=0}=2500$.  The signal-to-background ratio (the number of ev\-ents) 
collected in one month 
for the dimuons in J/$\psi$ and $\Upsilon$ mass regions with both particles in $|\eta|<0.8$ region are 2.75 (12600) and
0.52 (6000) for $dN_{ch}/d\eta|_{\eta=0}=5000$ and 4.5 (11600), 0.97 (6400) for
$dN_{ch}/d\eta|_{\eta=0}=2500$.
The background and reconstructed resonance numbers are in a mass interval  
$\pm \sigma$, where $\sigma$ is the mass resolution.

These quantities have been calculated for an integrated
luminosity of 0.5 nb$^{-1}$ assuming an average luminosity $\cal L$ = 
$ 4 \times 10^{26}$ cm$^{-2}$s$^{-1}$ and a machine efficiency  of 50\%. 
The expected statistics are large enough to allow further offline analysis for example 
in correlation with the centrality of the collision or the transverse momentum of the
resonance.

\section{Photon-tagged jet production}

Another important hard probe very sensitive to the initial conditions of the produced 
QCD-matter in heavy-ion collisions, is the QCD jet production. It is expected that 
final state in-medium interactions should reduce the energy of the jet partons  
(``jet quenching'') and result in medium-modified jet 
fragmentation~\cite{Baier:2000m}. Recent RHIC data on high-p$_T$ hadron
production~\cite{whitepapers} are consistent with jet quenching predictions. 
However full event-by-event reconstruction of jets in heavy ion collisions is rather  
complicated in the lower energy RHIC experiments. In CMS, large transverse momentum 
probes can be isolated experimentally from the soft particle background of the 
collision. In particular, full jet reconstruction and high-p$_T$ particle 
reconstruction in the high multiplicity environment of a Pb+Pb collision are
possible~\cite{D'Enterria:2007xr}. At the LHC, the production rates for jet pairs with 
transverse energy $E_T>50$ GeV are several orders of magnitude larger than at RHIC. 
Thus, high statistics systematic studies will be possible in a controlled perturbative 
regime, far beyond the limits of RHIC. The one of the important jet-related observables 
accessible to study in heavy ion collisions at CMS will be the $\gamma$-jet 
(and $Z$-jet) channel provides a very clean means to determine medium-modified parton 
fragmentation functions (FFs)~\cite{Arleo:2004xj}. Since the prompt $\gamma$ is not 
affected by final-state interactions, its transverse energy ($E_T^{\gamma}$) can be used as a 
proxy of the away-side parton energy ($E_T^{\rm jet}\approx E_T^{\gamma}$) before any 
jet quenching has taken place in the medium. The FF, i.e. the distribution of hadron 
momenta, $1/{N_{\rm jets}}\,dN/dz$, relative to that of the parent parton 
$E_T^{\rm jet}$, can be constructed using $z = p_T/E_T^{\gamma}$ or, similarly, 
$\xi=-\ln z=\ln (E_T^{\gamma}/p_{T})$, for all particles with momentum $p_T$ associated
 with the jet.

\begin{figure}[htb]
\includegraphics[width=7.7cm,height=6.5cm]{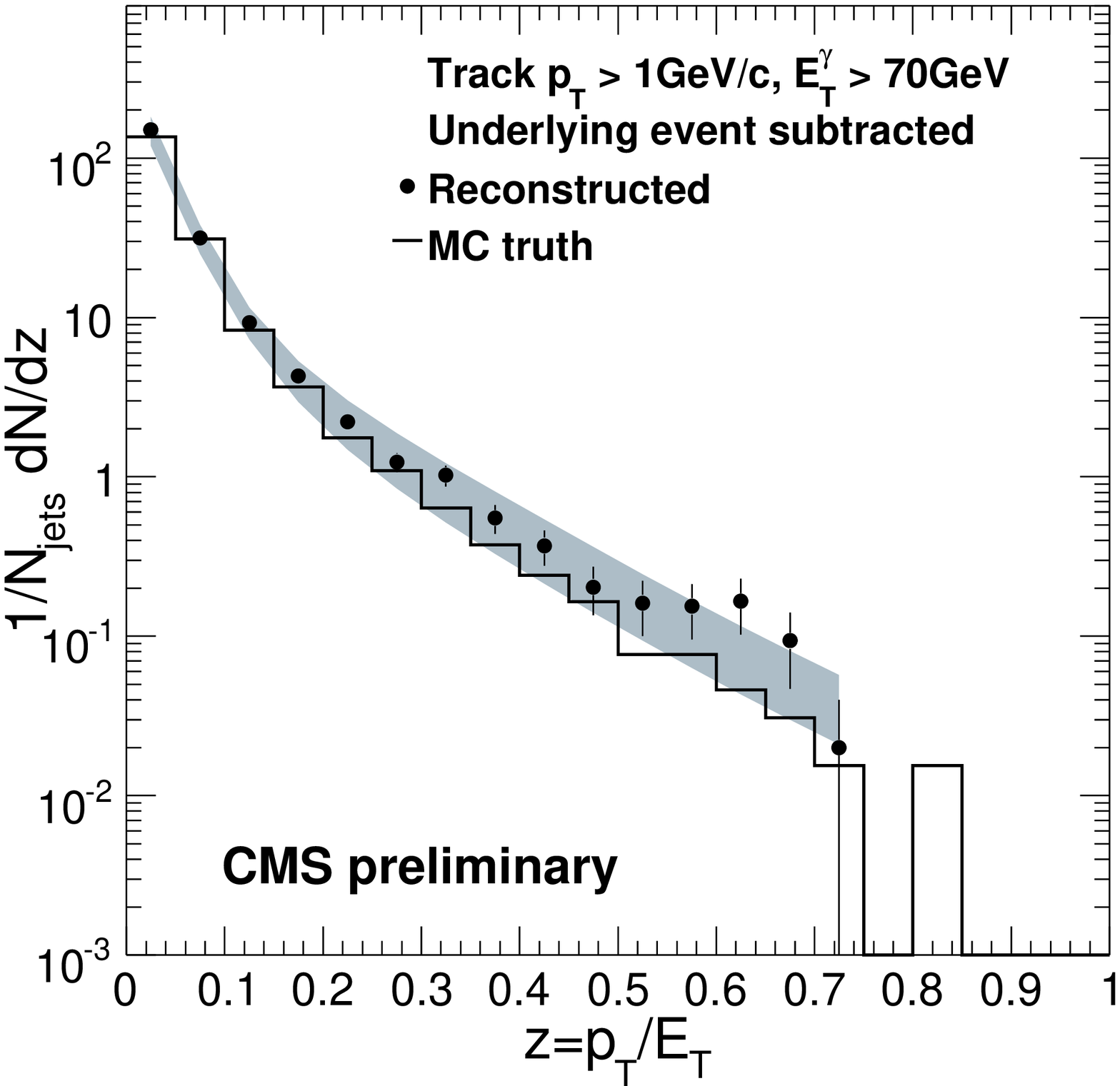}
\includegraphics[width=7.9cm,height=6.8cm]{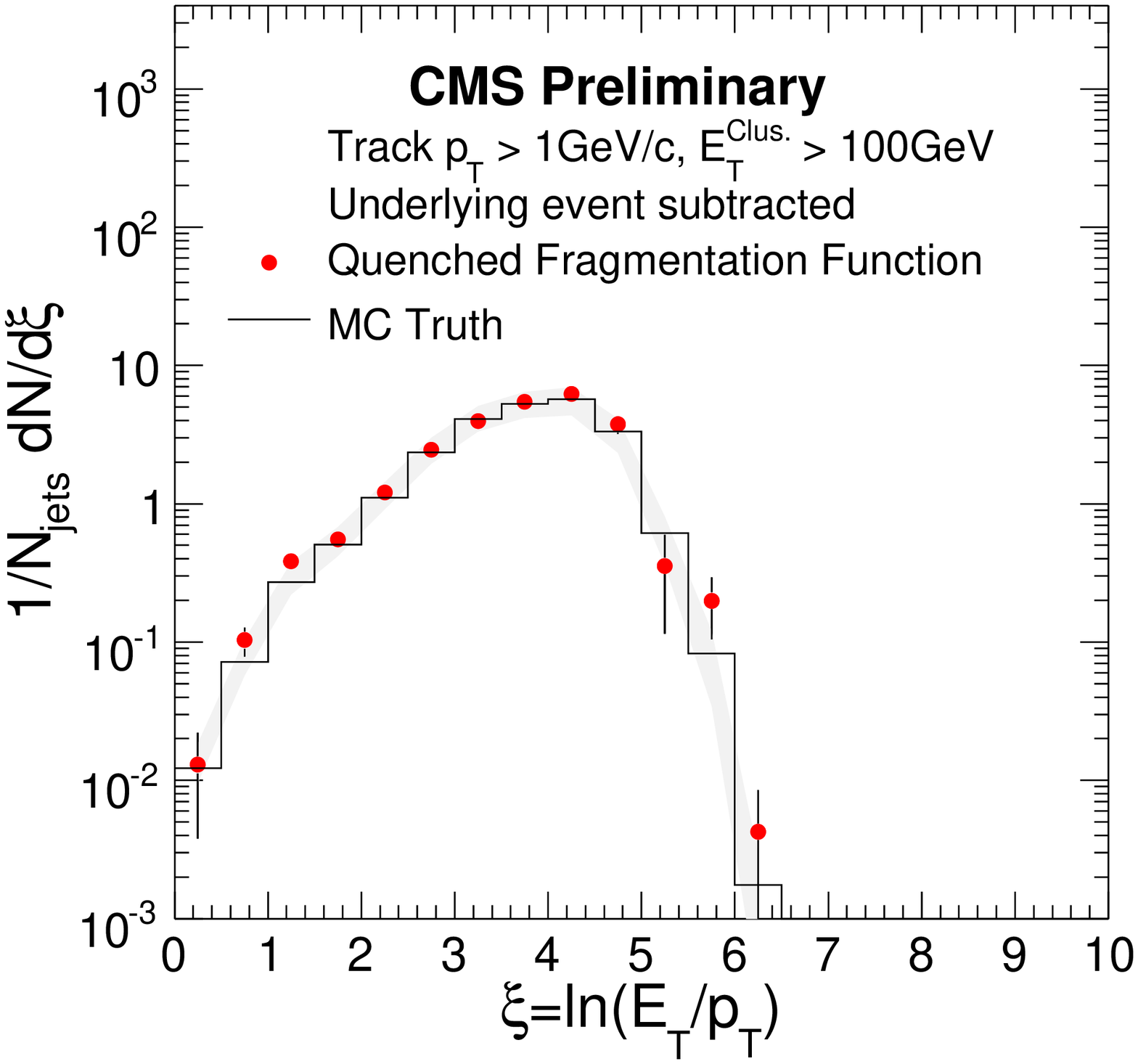}
\caption{Generated (histogram) and reconstructed (points) fragmentation functions 
as a function of $z$ (left) and $\xi$ (right) for quenched partons. Statistical errors 
correspond to an integrated luminosity of 0.5~nb$^{-1}$. The estimated systematic 
error is represented as the shaded band.}
\label{fig:mFFs}
\end{figure}

\begin{figure}[htbp]
\begin{center}
\includegraphics[width=8.2cm,height=7cm]{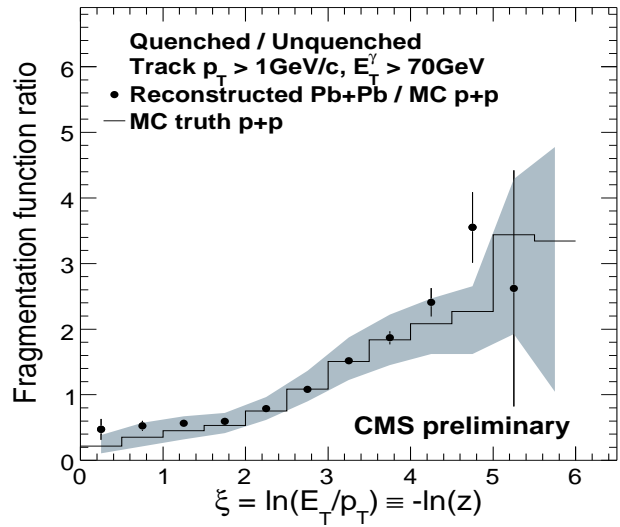}
\caption{Generated (histogram) and reconstructed (points) ratio of quenched (PYQUEN) 
and unquenched (PYTHIA) fragmentation functions as a function of $\xi$.}
\label{fig:quenunquen}
\end{center}
\end{figure}

Full CMS simulation-reconstruction studies of the $\gamma$-jet channel have been carried 
out~\cite{Loizides:2008pb}, where the isolated $\gamma$ is identified in ECAL 
($R_{isol}$~=~0.5), the away-side jet axis ($\Delta\phi_{\gamma-jet}>$~3~rad) is 
reconstructed in ECAL+HCAL, and the momenta of hadrons
around the jet-axis ($R_{jet}<$~0.5) are measured in the tracker. The event generators 
PYTHIA~\cite{pythia} (non-quenched partons) and PYQUEN~\cite{pyquen} (quenched partons) 
were used to simulate signal $\gamma$+jet events, and HYDJET~\cite{pyquen} was used to 
model the underlying (background) heavy ion event (also without or with jet quenching). 
For this study, the $10$\% most central Pb+Pb collisions were selected by the impact 
parameter of the lead nuclei, yielding an average mid-pseudorapidity density of about 
2400~(2200) charged particles in the quenched~(unquenched) HYDJET background. In total, 
4000 $\gamma$-jet events in the CMS acceptance for $E_T^{\gamma}>70$~GeV and 
$\mid{\eta^\gamma}\mid <2$ and about 
40000~(125000) QCD background events for the quenched~(unquenched) case are simulated.
This corresponds to the expected yields for one running year of Pb+Pb data taking with 
an integrated luminosity of 0.5~nb$^{-1}$. The working point for this analysis was set 
to 60\% signal efficiency, leading to a background rejection of about 96.5\%, and to a 
signal-to-background ratio of $4.5$ for $0-10$\%~central quenched Pb+Pb. 

The obtained FFs for photon-jet events with $E_T^{\gamma}>70$~GeV --- after subtraction of the 
underlying-event tracks using a R~=~0.5 cone transverse to the jet --- are shown in 
Fig.~\ref{fig:mFFs} for central quenched Pb+Pb collisions. Medium modified FFs are measurable 
with high significance (the systematic uncertainties being dominated by the low jet reconstruction 
efficiency for $E_T^{\rm jet}$~=~30--70~GeV) in the ranges $z<$~0.7 or 0.2~$<\xi<$~5.
The overall capability to measure the medium-induced modification of jet fragmentation functions 
in the $\gamma$+jet channel can be illustrated by comparing the fully reconstructed quenched 
fragmentation function to the unquenched MC truth distribution~(Fig.~\ref{fig:quenunquen}). 

\newpage

\section{Summary} 

With its large acceptance, nearly hermetic fine granularity hadronic and 
electromagnetic calorimetry, and good muon and tracking systems, CMS is an excellent 
device for the study of hard probes (such as quarkonia, jets, photons and high-p$_T$
hadrons) in heavy ion collisions at the LHC. 

\section{Acknowledgments} The author wishes to express the gratitude to the members 
of CMS Collaboration for providing the materials and the organizers of the Workshop 
``High-pT Physics at LHC'' 
for the warm welcome and the hospitality. The author also gratefully acknowledge 
support from Russian Foundation for Basic Research (grants No 08-02-91001 and No 
08-02-92496) and Grants of President of Russian Federation 
(No 1007.2008.2 and No 1456.2008.2).


\begin{thebibliography}{99}
\bibitem{D'Enterria:2007xr} D. d'Enterria (ed.) et al. (CMS Collaboration),   
\emph{J. Phys.} {\bf G 34} (2007) 2307.
\bibitem{ferenc} F. Sikler (CMS~Collaboration), in these Proceedings.
\bibitem{krisztian} K. Krajczar (CMS~Collaboration), in these Proceedings.
\bibitem{ptdr} A. de Roeck (ed.) et al.(CMS Collaboration),  
\emph{J. Phys.} {\bf G 34} (2007) 995.
\bibitem{Htdr} \emph{CMS HCAL Technical Design Report}, CERN/LHCC 97-31, 1997.
\bibitem{Mtdr} \emph{CMS MUON Technical Design Report}, CERN/LHCC 97-32, 1997.
\bibitem{Etdr} \emph{CMS ECAL Technical Design Report}, CERN/LHCC 97-33, 1997.
\bibitem{Ttdr} \emph{CMS Tracker Technical Design Report}, CERN/LHCC 98-6, 1998.
\bibitem{fwd_cms} M. Albrow et al. (CMS/TOTEM Collaborations),  
CERN-LHCC-2006-039/G-124, 2006. 
\bibitem{groland} G. Roland et al. (CMS Collaboration), \emph{J. Phys.} {\bf G 34} 
(2007) S733. 
\bibitem{Matsui:1986dk} T. Matsui and H. Satz, \emph{Phys. Lett.} {\bf B 178} (1986) 
416.
\bibitem{Abreu:1999qw} B.~Alessandro et al. (NA50 Collaboration), \emph{Eur. Phys. J.} 
\textbf{C 39} (2005) 335.
\bibitem{Abreu:2000ni} B.~Alessandro et al. (NA50 Collaboration), \emph{Eur. Phys. J.}  
\textbf{C 49} (2007) 559.
\bibitem{Adare:2006ns}  A.~Adare et al. (PHENIX Collaboration), \emph{Phys. Rev. Lett.} 
{\bf 98} (2007) 232301.
\bibitem{hijing} M.~Gyulassy and X.-N.~Wang, \emph{Comput. Phys. Commun.} {\bf 83} 
(1994) 307. 
\bibitem{Baier:2000m} R.~Baier, D.~Schiff and B.G.~Zakharov, \emph{Annual Rev. Nucl. 
Part. Sci.} {\bf 50} (2000) 37. 
\bibitem{whitepapers} \emph{RHIC White Papers}, \emph{Nucl. Phys.}  {\bf A 757} 
(2005) 28.
\bibitem{Arleo:2004xj} F. Arleo et al. \emph{JHEP} {\bf 0411} (2004) 009.
\bibitem{Loizides:2008pb} C. Loizides (CMS~Collaboration), {\tt arXiv:0804.3679}. 
\bibitem{pythia} T.~Sjostrand, S.~Mrenna and P.~Skands, \emph{JHEP} {\bf 0605} (2006) 
026.
\bibitem{pyquen} I.P.~Lokhtin and A.M.~Snigirev, \emph{Eur. Phys. J.} {\bf C 45} (2006) 
211.



\end{thebibliography}
\end{document}